The Galway Astronomical Stokes Polarimeter: An All-Stokes Optical Polarimeter with Ultra-High Time Resolution.

Patrick Collins[1], Gillian Kyne[1], David Lara[2], Michael Redfern[1], Andy Shearer[1], Brendan Sheehan[3]

**Abstract**

Many astronomical objects emit polarised light, which can give information both about their source mechanisms, and about (scattering) geometry in their source regions. To date (mostly) only the linearly polarised components of the emission have been observed in stellar sources. Observations have been constrained because of instrumental considerations to periods of excellent observing conditions, and to steady, slowly or periodically-varying sources. This leaves a whole range of interesting objects beyond the range of observation at present. The Galway Astronomical Stokes Polarimeter (GASP) has been developed to enable us to make observations on these very sources. GASP measures the four components of the Stokes Vector simultaneously over a broad wavelength range 400-800nm., with a time resolution of order microseconds given suitable detectors and a bright source - this is possible because the optical design contains no moving or modulating components. The initial design of GASP is presented and we include some preliminary observational results demonstrating that components of the Stokes vector can be measured to <1% in conditions of poor atmospheric stability. Issues of efficiency and stability are addressed. An analysis of suitable astronomical targets, demanding the unique properties of GASP, is also presented.

1. Introduction

1.1  Science

Optical polarisation is a valuable diagnostic tool in astrophysics, frequently enabling symmetries in source regions, magnetic field configurations, and magnetic field strengths to be investigated. Polarisation can be considered to be either intrinsic, that is coming from the object's source function, or extrinsic, in that the radiation is modified after emission primarily through scattering. Optical observations of astronomical objects which benefit from polarimetric observations include active galactic nuclei (Angel & Stockman 1980; Laor et al 1990; Agol & Blaes 1996; Goosman & Gaskell 2007), compact binaries (Patterson, 1994; Wickramasinghe & Meggitt 1985; Cropper 1990; Stockman et al 1992; West 1989; Katajainen et al 2007) and pulsars (Słowikowska et al 2009). In all these cases the emission comes from non-thermal sources - most likely the interaction of relativistic plasma with a strong magnetic field. Stellar scale binary systems normally show distinct stochastic and periodic variability; consequently, *time resolved* polarimetric observations are important. For GASP our main targets are isolated neutron stars, either pulsars or magnetars, and compact binary systems. Shearer et al (2010) give a review of the current state of optical polarimetric observations of these systems.

- Optical studies of pulsars have been limited by their intrinsic faintness. However, such studies are important because the high energy emission process is thought to be incoherent synchrotron or curvature radiation, and we expect this emission to be polarised. We can therefore expect to derive an important insight into the geometry of the emission process from optical observations. In the radio regime, where most pulsars have been discovered, the emission process is incoherent and complex making it more difficult to map radio polarisation directly to emission zone geometry. At photon energies above the radio it is only in the optical where we can measure all the Stokes parameters of an emitted beam to determine both linear and circular polarisation. Gamma-ray polarisation has been measured at low precision for one pulsar, the Crab (Dean et al 2008) - although it is likely that the polarised signal is coming from a nearby synchrotron knot and not the pulsar

1 Patrick Collins, Gillian Kyne, Michael Redfern, Andy Shearer
School of Physics, National University of Ireland, Galway, Ireland.
2 David Lara, Cambridge Consultants, Science Park, Milton Road, Cambridge CD4 0DW, UK
3 Brendan Sheehan, Tyndall National Institute, Lee Maltings, Cork, Ireland.
Corresponding author mike.redfern@nuigalway.ie  +353 (0)91 492490

itself (Hester et al 2002). No X-ray observations of the polarisation have been made although a future X-ray mission has been planned (McConnell et al, 2009). Optical polarimetric observations have been limited to the Crab pulsar (Smith et al 1988) and Słowikowska et al 2009) and with very low precision to PSR B0656+14 (Kern et al 2004) and PSR B0540-69. On the basis of the optical polarisation observed by Smith et al (1998), McDonald et al (2011) were able to determine, through an inverse mapping process, the location of the optical emission zone. Słowikowska et al's (2009) observations of the Crab pulsar showed that the degree of polarisation peaked, not with the peak optical emission, but in phase with the radio precursor. This observation has still to be explained. A similar link between optical and radio emission was observed by Shearer et al (2003) who saw a small increase in the optical flux during giant radio pulses (GRP) from the Crab pulsar. Observing the optical emission during GRP events was one of the main driving design factors behind GASP.

- Magnetars or Anomalous X-ray Pulsars are another class of neutron star, but here the dominant emission comes from X-rays. They are anomalous as their X-ray luminosity exceeds their spin-down energy. It is thought that the energy behind the emission comes from the decay of the star's exceptionally strong magnetic field, but precisely what the emission mechanism is remains a mystery. Optical observations and in particular polarimetric observations will be important to tie down the emission process. To date, no observations have been made of the optical polarisation from magnetars; see Shearer et al (2010) for a discussion of the possibilities.

- Optical observations of the brightest optical counterparts to Gamma-ray Bursts (such as GRB 090102) are within the brightness range of GASP targets, and can be polarised. Although GASP's field of view means that the actual burst is highly unlikely to be observed the afterglow can be. Observations of periodic polarised behaviour would yield hugely important clues to their nature through an understanding of the geometry of the emission zone.

- Accreting compact binary systems are complex objects with matter flowing from a secondary star onto a compact primary. This process normally generates an accretion disk and associated jet. Near the inner part of the accretion disk and at the base of the jet the emission process is likely to be synchrotron radiation which will be both linearly and circularly polarised. One class of accreting compact binary systems is that of the cataclysmic variables (CVs), which occur when the compact primary's magnetic field is sufficiently strong to disrupt any accretion disk. These are classified as polars, when the rotation period of the primary is synchronised with the orbital period of the secondary star, or as intermediate polars when the periods are asynchronous. The inflowing material is fully ionised as it approaches the compact object's surface, producing cyclotron emission; this we expect to be highly polarised - showing both linear and circular polarisation (Megitt & Wickramasinghe 1982). From the polarisation signal we can determine the geometry of the jet's emission zone as well as that of the inner accretion disk (Antonucci 1982).

Table 1 gives a list of possible targets for GASP. These are mostly periodic sources, which may also show aperiodic behaviour. However, the brightest optical counterparts to Gamma-ray bursts (such as GRB 090102) are also included because they would be within the brightness range of GASP targets, and can be polarised. Observations of periodic behaviour would yield hugely important clues to their nature. Periodic polarisation variation has been measured for the Crab nebula pulsar only – indicated as (P) in the table. Otherwise pulsar polarisation measurements are aperiodic (A).

| Target[A] | Time Scale (s) | Magnitude (V) | GASP Flux[B] counts/sec/channel | | Object Polarisation[C] (%) | Polarisation sensitivity[D] requirement (%) | | Exposure Time to 5% polarisation (s) | | Ref |
|---|---|---|---|---|---|---|---|---|---|---|
| **Pulsars** | | | 4m | 10m | | Object | Total | 4m | 10m | |
| Crab | 0.033 | 16.8 | 2800 | 17100 | 16 (P) | 0.1 | 0.1 | 6,000 | 1000 | 1,2 |
| Vela | 0.089 | 24 | 3.7 | 22.5 | 9.4+/-4 (A) | 1 | 0.03 | >12 hours | 13,800 | 3,4 |
| B0540-60 | 0.050 | 23 | 9.3 | 56.6 | <15(P) 5 (A) | 1 | 0.07 | 10,000 | 1,700 | 3,5,6 |
| B0656+14 | 0.385 | 25 | 1.5 | 9.0 | 0-100 | 2 | 0.03 | >12 hours | 15,000 | 7 |
| **Magnetars** | | | | | | | | | | |
| 4U 0142+61 | 8.69 | 25-26 | 0.8 | 5.1 | Not known | 5 | 0.03 | >12 hours | 15,000 | 8 |
| 1E 1841-045 | 11.78 | 25.3 (I) | 1.5 | 9.4 | Not known | 10 | 0.05 | >12 hours | 28,000[E] | 9 |
| **Gamma Ray Bursts** | | | | | | | | | | |
| GRB 090102 | 1 | 13(R) | 92,000 | 564,000 | 1% | 1 | 1 | 55 | 9[E] | 10 |
| **Close Binary Systems** | | | | | | | | | | |
| RX J0649.8-0737 (polar) | 15811 | 17.2 | 1940 | 11,800 | Not known | 0.1 | 0.09 | 6,200 | 1,030 | 11 |
| RXJ1914.4+2456 (intermediate polar) | 569 | 19.7 | 194 | 1,180 | Not known | 0.1 | 0.06 | 14,000 | 2,300 | 12 |

**Table 1:** Expected degree of polarisation, fluxes, and minimum exposure time for detecting 5% polarisation against normal sky background and 1" seeing/aperture.

[1] Słowikowska et al 2009, [2] Smith et al 1988, [3] Wagner & Seifert 2000, [4] Mignani et al 2007, [5] [Middleditch et al. 1987, [6] Chanan & Helfand 1990, [7] Kern et al 2003 [8], Hulleman et al 2000 [9] Dhillon et al 2010 [10], McConnell et al 2009, [11] Motch et al 1998, [12] Strohmayer 2004

Optical magnitudes are V band except where otherwise indicated. Flux counts and exposure times to determine polarisation to within 5% are indicated for observations using GASP mounted on a 4-m class and a 10-m class telescope. We note that with the exception of the brighter targets, all of these targets are too faint for a 4-m class telescope.

**Notes on Table 1:**

For faint targets where the flux is significantly less than the sky flux the exposure time is the time to detect the sky background to the required sensitivity.

A. Four main groups of targets have been identified. For the first two groups, pulsars and magnetars, the time scale given is the rotation period. For the Gamma-Ray bursts a one second integration period has been used. For the binary systems the orbital period of the secondary has been specified. We note that the rotation period of the primary is likely to be significantly less than this.
B. The GASP flux is based upon the measured throughput of GASP and assuming 70% detector quantum efficiency and 95% total reflectivity of the telescope mirrors.
C. The given polarisation is either for the pulsed signal (**P**) or the time-averaged signal (**A**). In terms of understanding the physics of the objects it is usually the pulsed signal which is of primary interest.
D. The figure given here shows the sensitivity required for the science case and the actual object polarisation assuming an unpolarised sky background. A V-band sky background of 21.5magnitudes/arcsec$^2$ has been assumed. All the figures refer to linear polarisation. The total column indicates the accuracy to which the polarimeter must be calibrated to achieve the sensitivity requirement. The sensitivity requirement would be similar for linear and circular, however, with the caveat that to date circular polarisation has been observed in polars and intermediate polars, but not for any of the other objects.
E. 1E 1841-045 and GRB exposure times have been based upon I and R band sensitivity and sky background respectively.

## 1.2 This Paper

In this paper we describe the development of an optical instrument which is capable of characterizing an astronomical signal by *simultaneous* measurement of its four Stokes parameters with a precision and temporal resolution for stochastically-varying signals limited only by the optical flux and by the speed, efficiency and stability of its detectors. Using suitable detectors, such an instrument would be ideally suited to measure the optical polarisation from close binary systems and isolated neutron stars, free of timing noise and capable of measuring both stochastic and periodic variability on times scales ranging from milliseconds (or better, given very high pulsed fluxes) to hours.

We call our demonstration of such a polarimeter the Galway Astronomical Stokes Polarimeter (GASP).

## 1.3 Polarimetry Principles

The polarisation state of a beam is defined by the Stokes vector **S**,

$$S = [I, Q, U, V]^T = [\langle I_x + I_y\rangle, \langle I_x - I_y\rangle, \langle I_{+45°} - I_{-45°}\rangle, \langle I_{Rc} - I_{Lc}\rangle]^T \quad (1)$$

where $I_x$, $I_y$, $I_{+45}$, and $I_{-45}$ are the intensities of the linear polarisation components of the beam in the respective directions, and $I_{Rc}$ and $I_{Lc}$ are the intensities of the right and left hand circularly polarised components, respectively. T denotes transposition.

In order to determine these four Stokes components, at least four independent measurements must be made - i.e.. the beam must be divided into four components and the intensity of each component must be measured. These intensities represent the projection of the original Stokes vector onto each of the components measured. It is convenient to write the four intensities as a column vector $I = (i_1, i_2, i_3, i_4)^T$, which is linearly related to the original Stokes vector **S** by the system matrix **A** as follows:

$$I = \mathbf{A}.S \quad (2)$$

The system matrix, **A,** is, in general, an Nx4 matrix, with each row corresponding to the polarisation eigenvector of each of the components chosen in the design of the polarimeter. In this paper our design makes N=4, and from now on, we will regard **A** as a 4x4 matrix of real numbers. From Eq. (2), it is clear that S can only be uniquely determined from I if the matrix **A** can be inverted – which is to say that each of the components of S must affect I. The exact components of **A** are determined by design and then determined through careful experimental calibration.

If **A** can be inverted, then S can straightforwardly be determined from I as

S= **A**$^{-1}$.I (3)

The division of the original input beam into four output beams, incorporating the different polarisation projections, can be done by many methods, but they mainly fall into three classes, namely:

Division Of Time Polarimeters (DOTP) - in which the components of the intensity vector I are determined sequentially, by using rotating or modulating polarisation elements. This is the most common type of polarimeter– seen, for example, in Optima (Straubmeier, Kanbach and Schrey 2003). An intrinsic disadvantage of the DOTP methodology is that the sampling (modulation) frequency imposes a high frequency cutoff in the estimation of randomly polarised signals – nevertheless, even high speed periodic variability can be determined phase by phase through averaging over many signal cycles. The speed of the system is always limited by how quickly the modulation of the polarisation elements can be performed. However, polarimeters of this type are relatively easy to implement, and if the signals of interest tend to fluctuate slowly very sensitive measurements can be made. For these reasons, DOTP imaging polarimeters are the preferred solution for current very large and future extremely large telescopes such as SPHERE-ZIMPOL on the VLT (Roelfsema et al 2011).

Division Of Wavefront Polarimeters (DOWP) - in which at least four fractional areas of the wavefront, in, say, a collimated beam are separated and analyzed separately to determine the Stokes components. This method has been used for imaging polarimetry in Earth Observations studies, etc (Duggin et al 1990 and Whitehead et al 1990).

Division Of Amplitude Polarimeters (DOAP) - in which the amplitude of the whole wavefront, in, say, a collimated beam is divided into at least four components from which the Stokes components can be determined. The polarimeter described in this paper (GASP) utilises the DOAP principle. Examples of DOAP in optics can be found in Azzam (1982), Compain & Drevillon (1998), Lara and Dainty (2005), Lara and Dainty (2006), and Lara and Paterson (2009). Examples of DOAP in astronomical instrumentation include: ExPo (imaging polarimetry, Rodenhuis et al 2012), the HARPS polarimeter (spectroscopy, Snik et al 2011), HiCIAO (imaging polarimetry, Suzuki et al 2009), and PlanetPol (spectroscopy, Hough et al 2006). Some of these (ExPo, PlanetPol) combine DOAP and DOTP, which can be shown to have better sensitivity and accuracy than either one of these two methods alone (Semel et al 1993, Keller 1996). However, the use of the DOAP methodology alone does not, in principle, generate a high frequency cutoff since the four component beams can be detected simultaneously (although this was not the case in the design of Compain and Drevillon (1998) from which the design of GASP is derived). In the case where very high time resolution is a critical consideration, the DOAP methodology, as used in GASP, provides the best solution.

## 2. Optical Design of GASP

The Galway Astronomical Stokes Polarimeter (GASP) described in this paper, works by splitting the beam (S) into two components, and then using a suitably oriented polarising beam splitter (PBS) in each of the two beams, dividing them again into two orthogonal linearly polarised beams (4 beams in all), of intensities $i_1$-$i_4$. The ratios of intensities $i_1$ and $i_2$ carry information about Stokes parameters Q and U. An anachromatic (or at least very wide band) quarter wave phase retardation (turning circular into linear polarisation) and additional diattenuation is introduced into one of the beams which the second PBS can then examine - the ratios of $i_3$ and $i_4$ therefore carry information about the Stokes parameters Q and V, as will be explained later. Information about I is carried by the sum $i_1 + i_2 + i_3 + i_4$.

The design of GASP is based on a DOAP concept using partial reflection from the uncoated end surfaces of a rhomb-type prism, similar to a Fresnel rhomb. Two internal reflections within the rhomb introduce a relative phase delay between the p and s waves of the transmitted light, which is dependent upon the refractive index of the glass and the angle of total internal reflection. For glasses of typical refractive index (1.4-1.6) and suitable total internal reflection angles the relative phase delay can be 45° at each reflection – producing an effective quarter wave retardation overall. Hence, the rhomb can be used to convert linear to circular polarisation and vice versa. The principle of the design in GASP makes use of additional polarisation-dependent transmission/reflection introduced when the beam enters and exits the prism at steep angles. This was first reported by Compain & Drevillon (1998) in their paper "Broadband division of amplitude polarimeter based on uncoated prisms". In this paper we make use of the results reported by Compain & Drevillon (1998) and we optimise their device for astronomical observations.

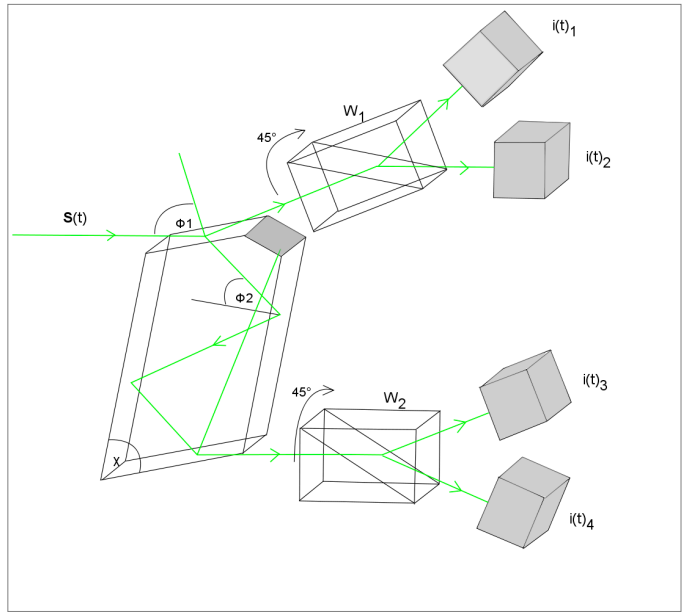

**Fig. 1** Principle of a visible-near-infrared DOAP with an uncoated dielectric prism (diagram after Compain & Drevillon (1998))

The entrant collimated light beam S(t) is separated into two beams: R(t)-reflected, and T(t)-transmitted. The expected intensities of S(t) and R(t) are arranged to be approximately equal when fully unpolarised light is incident on the prism. We denote by $\phi_1$ the angle of incidence on the uncoated face of prism, and by $\chi$, the internal angle that defines the shape of the parallel sided prism. R(t) and S(t) are each subsequently divided into two by two polarising prisms (Wollaston prisms, W1 and W2, in this case) oriented at 45° to the plane of incidence on the prism. The internally transmitted beam T(t) undergoes a quarter wave uniaxial phase delay created by two total internal reflections at angle $\phi_2$ (determined by the geometry and the refractive index of the glass) inside the prism, and is then partially transmitted and partially reflected at the bottom face of the prism.

The Stokes vector can then be determined from the intensities $i(t)_1$, $i(t)_2$, $i(t)_3$ and $i(t)_4$. In the original design (Compain and Drevillon (1998)) a small fraction (~20%) of the light was inevitably reflected at the second partial transmission and was absorbed by a blackened vertex of the prism as shown in Figure 1.

From Table 1 we see that the potential targets are faint; consequently, GASP is required to operate as efficiently as possible – precluding therefore the use of polarising filters (see, for example, Smith at al 1992, Słowikowska et al 2009, and Kern et al 2003) - and over a wide wavelength range (400-800 nm is a practical limit, set by detector spectral response). Therefore, in GASP we require that the quarter wave phase retardation be <0.1° over the wavelength range 400-800nm. Figure 2 (a) below, shows how this may be achieved. It is a graph of retardance versus angle of internal reflection for glasses of various refractive indices, similar to Figure 2 in Compain and Drevillon's (1998) paper. The prism designed for Compain and Drevillon's (1998) used region A in figure 2(a) to ensure it gave the required retardance. We note that region B of figure 2(a) could also yield the required $45^0$ retardance, but with less dispersion, and this is where we work.

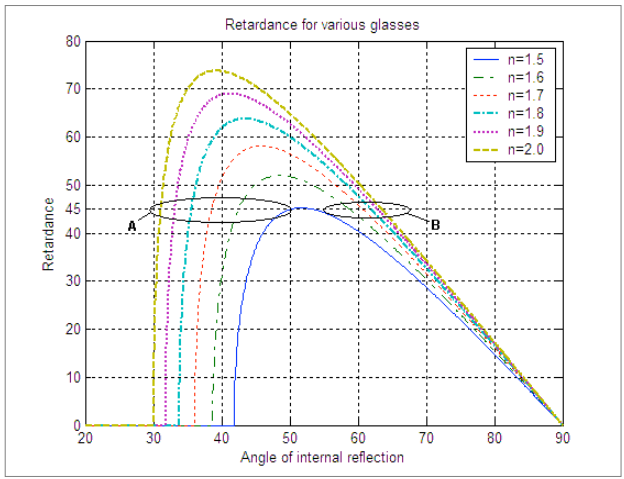 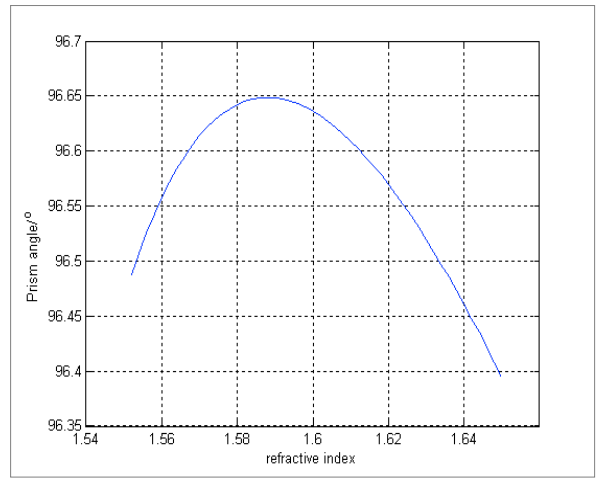

(a)  (b)

**Fig. 2(a)** Design requirement for the prism angle vs. wavelength to meet the twin conditions of equal transmitted and reflected beams and a precise quarter wavelength internal retardation.
**Fig. 2(b)** The retardance due to total internal reflection as a function of angle for various refractive index glasses.

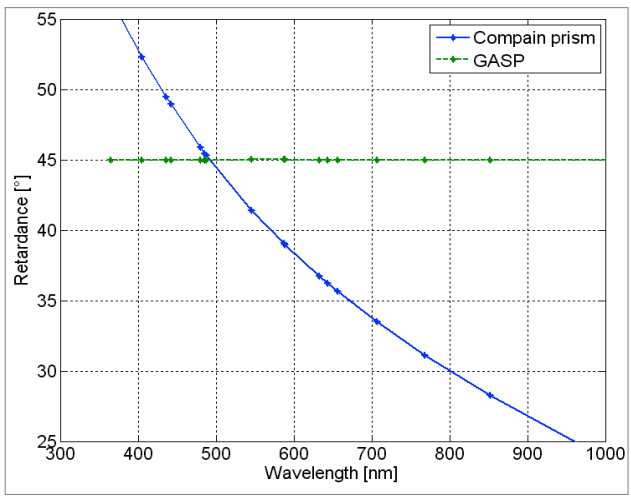 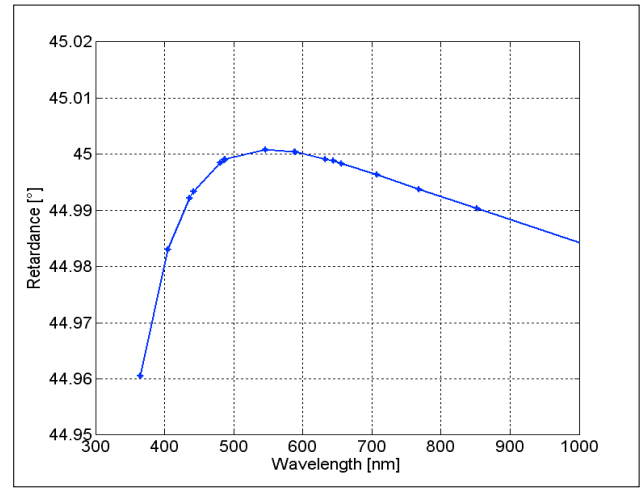

(a)  (b)

**Fig. 3(a)** A plot of retardance versus wavelength for the GASP prism and the Compain and Drevillon original RBS. Here, the retardance values of both prisms are compared and the improvement of the achromaticity of GASP from the new prism angle.
**Fig 3(b)** a magnified view of the retardance for the GASP prism where we can see that it is tuned for 589nm.

With the requirement for R(t) and T(t) beams to be approximately equal and a retardance of a quarter wave we end up with figure 2(b) where we can see that, ideally, a low dispersion glass with an refractive index around 1.59 could give us an achromatic retardance. When the reflected and transmitted intensities are approximately equal for incident unpolarised light, This gives a good condition number for inversion of the system matrix A (Eq. 2) – which allows it to be inverted without generating excessive errors – and hence is used for the optimization of the prism design. It is also common practice to optimize the polarimeter using the condition number of the system matrix **A** (Tyo 2002), which maximizes the volume of a tetrahedron generated by the four vectors in the Poincaré sphere, and distributes the noise more or less evenly among the four Stokes parameters.

Commercially available glasses of high transparency can be found to with a refractive index of ~1.59 and with a low dispersion where Schott N-SK5 was actually used. Figure 3(a) and 3(b) gives an overall performance of the final GASP prism after its enhancement. We note that this glass gives us a large bandwidth with a retardance error <<1% over the required spectral range for GASP of 400nm to 800 nm. The eventual prism angle chosen was 96.62°. For comparison, the best super-achromatic quarter wave retarders have a retardance error of 3.6° in the wavelength range 400-800 nm (Frecker et al 1976).

In Figure 1, it can be seen that in the original Compain & Drevillon (1998) design there was an **inevitable** ~20% of the light reflected from the second face of the rhomb-type prism – this arises in the optimization of their prism design for polarimetric efficiency. In order to absorb this light, which might otherwise result in stray reflections, in their design one vertex of the prism was removed and the resulting surface roughened and blackened. In the design of GASP an additional "extractor" prism, was designed to produce transmission of 80% of the stray light into an external beam parallel to R(t) to be used for finding, guiding, and/or photometry – but not to determine components of the original Stokes vector. The final ~4% waste light is also absorbed by a blackened ground surface. Efficiency could be increased by recording also the orthogonal states of polarisation in the extractor arm, producing a 6x4 matrix of intensities. The Moore-Penrose pseudo-inverse can then be used to determine the input Stokes vector (del Toro Iniesta & Collados (2000)).

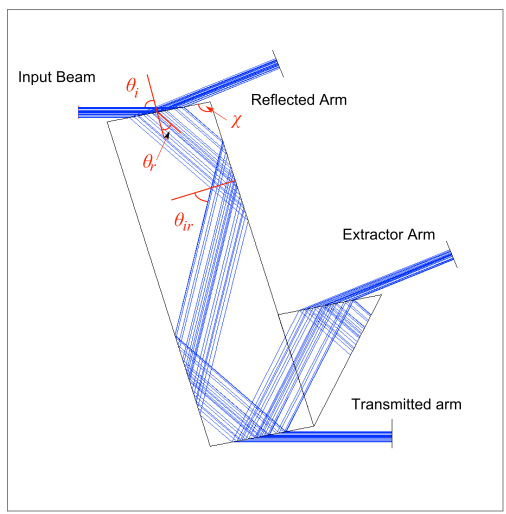

**Fig. 4** Final layout of the retarding beam splitter and the extractor prism used in GASP showing the input, reflected, transmitted and extracted beams

The general analytical form of the matrix **A** has been derived in Compain & Drevillon (1998), and is given below, using the same notation as in their paper. A full derivation of **A** is beyond the scope of this paper, which is intended to illustrate the adaption of the Compain & Drevillon original design to astronomical Stokes polarimetry.

$$A = -RT \frac{1}{2} \begin{bmatrix} 1 & -\cos(2\psi_r) & \sin(2\psi_r)\cos(\Delta_r) & \sin(2\psi_r)\sin(\Delta_r) \\ 1 & -\cos(2\psi_r) & -\sin(2\psi_r)\cos(\Delta_r) & -\sin(2\psi_r)\sin(\Delta_r) \\ 1 & -\cos(2\psi_t) & \sin(2\psi_t)\cos(\Delta_t) & \sin(2\psi_t)\sin(\Delta_t) \\ 1 & -\cos(2\psi_t) & -\sin(2\psi_t)\cos(\Delta_t) & -\sin(2\psi_t)\sin(\Delta_t) \end{bmatrix}$$

Where

$$RT = \begin{bmatrix} R & 0 & 0 & 0 \\ 0 & R & 0 & 0 \\ 0 & 0 & T & 0 \\ 0 & 0 & 0 & T \end{bmatrix}$$

The values of the ellipsometric angles, $\psi$ and $\Delta$, depend on the optical properties of the glass prism and also of the angles of incidence on the various optical surfaces – as defined in Figure 4, above.

For the glass chosen (O'Hara S-BSL7), these values are given below in Table 2.

| Prism Geometry (degrees) | Glass Properties | Polarimetric properties | |
|---|---|---|---|
| | | Reflection | Transmission |
| $\varphi_1$=79.25 | $\lambda$=589nm | R=0.376 | T=0.377 |
| $\varphi_2$=58.43 | n=1.589 | $\Delta_r$=180 | $\Delta_t$=90 |
| $\chi$=96.62 | $\alpha$ =0.2186 | $\psi_r$=31.43 | $\psi_t$=60.38 |

**Table 2** Geometrical and Ellipsometric Properties of the GASP rhomb-type prism

Substituting these values into the general form of the **A** matrix produces a theoretical (optimized) system matrix for the prism alone – not including effects due to extra optical components such as turning mirrors, etc., these are considered in section 3, below :

| 0.1881 | -0.0858 | 0.1674 | 0.0000 |
|---|---|---|---|
| 0.1881 | -0.0858 | -0.1674 | 0.0000 |
| 0.1887 | 0.0965 | 0.0000 | 0.1622 |
| 0.1887 | 0.0965 | 0.0000 | -0.1622 |

Various optical designs of GASP have been explored, in which the four beams $i_1$-$i_4$ have been detected in various ways. These include a single imaging detector design, a double imaging detector design, and a multiple non-imaging detector design (using an array of photon counting avalanche photodiodes). These different solutions are appropriate to different astronomical targets, because whilst the imaging solutions offer high efficiency and the ability to perform polarimetry on point or extended sources in complex field regions, under varying atmospheric seeing conditions, they have limited time resolution. For the highest time resolution photon counting detectors are required. However, whilst these considerations are peripheral to the optical design and performance of GASP, detector limitations (noise and stability) ultimately limit its performance. Figure 5 shows the optical layout for a version which uses one imaging camera (either CCD, EMCCD or sCMOS cameras – for references to these types of cameras see, for example, Andor Technology (http://www.andor.com/camera.aspx) as the detectors for the primary (polarimetry) and secondary (extracted, guiding) beams.

**Choice of detector for single-imaging-system**

GASP as demonstrated uses a single DV887-BV iXon camera system from Andor Technology. This camera has a Low Light Level CCD or L3CCD sensor; CCD97 from E2V technologies. This sensor is back illuminated thus providing an improved spectral response and peak quantum efficiency of ≈ 90% at 575 nm. It comes equipped with two readout modes – conventional and electron multiplication. Electron multiplication permits the amplification of low signal levels above the noise floor of the readout electronics and the original photon count can be extracted through the use of thresholding schemes (Basden et al., 2003). Table 2 summarises the DV887-BV iXon camera specifications.

| Active Pixels, Size | 512 x 512, 16 μm$^2$ | Binning modes | 1 x 1, 2 x 2, 4 x 4 |
|---|---|---|---|
| Readout pixel Rates | 1, 3, 5, 10 MHz | Peak QE (@ 575nm) | 92.5 % |
| Frame Rates | 31 → 400 fps | Dark Current (@ -90∘C) | 0.0035e$^-$/pixel/sec |

**Table 2: DV887-BV iXon Specifications**

The use of a frame readout detector, such as the iXon CCD, limits the time resolution to 2.5 ms, which will not allow measurement of the fine structure in the periodic behaviour which the Crab, for example, is known to exhibit (Słowikowska et al 2009). This is not a problem in the case of non-periodic signals where signal to noise considerations limit the time resolution to be much worse than this. However, the use of a single detector for all four beams eliminates the issues attributed to differential gain variations, as all data is acquired through the same readout electronics and thus also have the same noise characteristics. This is important as any variations in detector readout noise or gain between the four beams can result in significant errors in the polarimetry (Keller, 1996). For these reasons a single-detector imaging version was used to demonstrate GASP in the results below, and figure 5 shows the optical arrangement which was used to demonstrate GASP. Most astronomical targets have a low degree of polarisation so the four output beams will be of approximately equal intensity, therefore signal to noise will also be approximately equal in each channel, and detector non-linearity will not alter the intensity ratios.

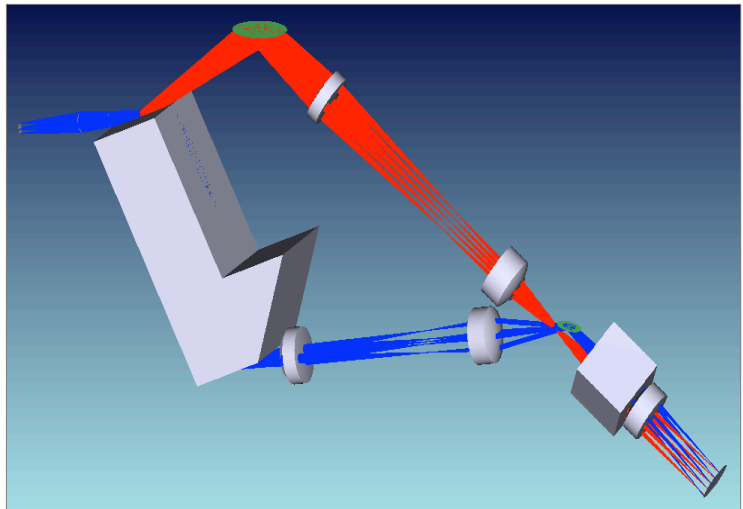

**Fig. 5** Optical design for a single-imaging-detector version of GASP (based upon Figure 4, above). Collimated light beam S(t) is split into two beams R(t) and T(t) by the retarding beam splitter, which are then incident upon two Wollaston prisms, each oriented at 45$^0$ to the plane of the original reflection. The resulting four beams are refocused upon one imaging detector using a 2-mirror solution to capture both beams. A third beam (~16 % of the total intensity) can be extracted and refocused on a further imaging detector. This can be used for guiding during observations for a non-imaging system.

The use of an imaging detector which reads out the data in frames precludes the achievement of very high time resolution, and previous very high time resolution instruments – such as Optima (described in Słowikowska et al 2009) – have used non-imaging photon counting avalanche photodiodes to achieve the necessary time resolution. A non-imaging version of GASP has been designed, as shown in Figure 6, below. Ancillary optical components, needed for calibration of the instrument A-matrix are common to imaging and non-imaging versions of GASP. Dispersion in the Wollaston prism produces chromatic distortion of the stellar images and makes accurate, very wide band, photometry difficult in the optical layout shown in Figure 5. This effect is reduced by using a low divergence angle Wollaston prism and by limiting the bandwidth to ~200nm by the use of a standard astronomical filter in the input beam. For simplicity, the design layout and other considerations of a non-imaging version will be described in this section.

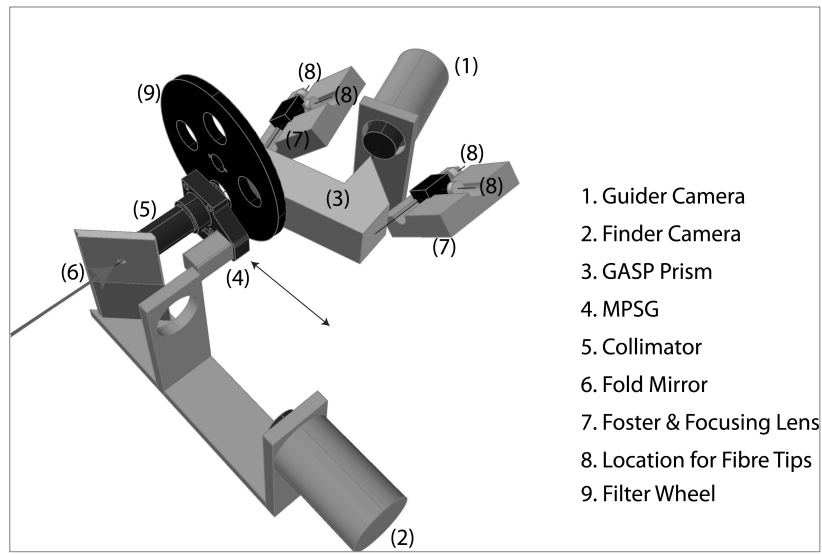

**Fig. 6** Optical Layout of a non-imaging (8 detectors) version of GASP

This (non-imaging, potentially extremely high time resolution) version of GASP uses an array of eight fibre-fed single photon counting avalanche photodiodes, arranged as four pairs in four bundles with fibre tips separated by 250 microns (one pair for each of $i_1$-$i_4$). Incoming light is focused in the plane of a small aperture (forming a field stop) in a diagonal mirror (6) used to form an off-axis image in the finder camera (2). The primary beam is collimated (5) to a beam diameter of ~4 mm, and passes through calibration and/or filtering components (4, 9). It then impinges on the top face of the GASP prism (3) at grazing incidence, forming two collimated beams by partial reflection at the top face and by partial transmission followed by two total internal reflections and partial transmission at the bottom face. These beams enter Foster polarising prisms (black components of 7) oriented with their two axes at $45^0$ to the plane of the top face of the prism. Foster prisms – rather than Wollaston prisms – are used in this case because their exit beams emerge at a right angle, which is mechanically convenient, and because broadband images are therefore not distorted by dispersion – as would be the case with Wollaston prisms. This enables the non-imaging design to use a very wide bandwidth (400-800nm, essentially determined by the detector responsivity) – necessary to observe faint sources on short timescales. Images emerging from the Foster prisms are focused on to four fibre pairs (8) collecting on-axis (target) and off-axis (sky background) light. Light from the extracted beam (~16%) is focused on to a guider camera (1) which is used for fine guiding, because it is crucial to keep the target centred on the on-axis fibre tips. The actual component values (focal lengths of collimating and focusing lenses, collimated beam diameter) depend upon at which particular telescope GASP has been designed to be used.

Calibration of GASP is performed in two stages; prior to mounting on the telescope, and after mounting. For the purpose of calibration, a linear polariser and a broadband quarter wave plate (4) are introduced in to the beam. These make up the GASP Polarisation State Generator (PSG). The linear polariser can be rotated, so that in combination with the quarter wave retarder, sufficient linearly independent polarisation states can be generated for calibration. Prior to mounting on the telescope an external source of unpolarised light in a converging beam with the same focus and focal ratio as the telescope can be mounted externally to the GASP enclosure. This is used to calibrate matrix **A**. After mounting, light from a bright unpolarised star is used for calibration. Alternatively, the finder camera can be replaced by a variable polarisation state generator, producing polarised light which can be introduced into the beam by a moving diagonal flat, replacing the diagonal mirror (6). This will allow full **A**-matrix calibration to be performed after mounting on the telescope. Introducing standard (broadband, unpolarised, correct focus and convergence) light via a partially-polarising diagonal flat with a metallic reflecting surface worsens but does not destroy the validity of the calibration.

A second filter wheel (part of 9), carries standard colour filters, for example g' r' and i', which can be introduced in to the beam, for calibration and coloured observations.

Photon counts from the eight counters are individually time-tagged to GPS time with a resolution of 250 ns. This enables GASP to investigate polarisation variations in the microsecond time regime, and to synchronize observations with those from other instruments or observatories, such as radio or other optical telescopes.

## 3. Calibration of GASP

Calibration of a complete Stokes polarimeter requires that the output intensity vector I be measured using at least four linearly independent polarisation. A typical example would be unpolarised light passing through one of three linear polarisers at 0°, 45° and 90° and an additional circular polariser. In principle, the selection of these four polarisation states depends on the final task of the instrument. However, for astronomical observations, S is unlikely to be near to 100% linear or circular polarisation, and is more likely to be nearly unpolarised.

The experimental estimation of the system matrix **A** was carried out with a pseudo-inverse computation. A set of at least 4 linearly independent input Stokes vectors were organized into a **PSG** (polarisation state generator) matrix. The measured intensities were correspondingly ordered in matrix **I**. To obtain the system matrix the intensity matrix was multiplied by the inverse of the analytically calculated **PSG** matrix.

**A= I.PSG$^{-1}$** (4)

To compensate for instrumental errors incurred in the generation of the Stokes vectors, the number of illumination states was over-determined. The pseudo-inverse of **PSG** was then be used to estimate, by a least squares method, the final system matrix **A**.

One simple way to generate the illumination Stokes vectors in the **PSG** was by the use of a rotating linear and removable quarter wave plate placed in the input beam. Firstly a polariser was rotated from 0° to 170° in steps of 10° to yield linear polarisations. A quarter wave plate was then inserted after the polariser and once again the polariser was rotated. This generated elliptically polarised states. The combination of these 36 Stokes vectors over-determined the **PSG**. It should be noted that the angles of the polariser and the fast axis of the quarter wave plate relative to the instrument's axis must be determined first of all at the design stage, because the **PSG** matrix for the estimation of **A** is calculated analytically. This type of calibration is subject to errors because of imperfections in the calibration component properties or orientation. This issue is addressed in section 6.1, below. The **A** matrix given above is that for the prism alone. Of course, the optical system contains other components which will feed into the matrix values, particularly non-common-path components such as the two turning mirrors at different angles in the reflected and transmitted beams, which will create cross-talk between the matrix components. The normalized matrices given below include the effects of these components,

| 1 | -0.5132 | -0.8217 | -0.2477 |
|---|---------|---------|---------|
| 1 | -0.5132 | 0.8217  | 0.2477  |
| 1 | 0.4388  | 0.4134  | -0.7982 |
| 1 | 0.4388  | -0.4134 | 0.7982  |

Theoretical GASP System Matrix at 650nm

| 1 | -0.4138 | -0.7446 | -0.4550 |
|---|---------|---------|---------|
| 1 | -0.4712 | 0.7891  | 0.4759  |
| 1 | 0.5634  | 0.6326  | -0.5139 |
| 1 | 0.4775  | -0.6726 | 0.5220  |

Experimental GASP System Matrix at 650nm
The matrixes are similar – differences are due to an inadequate knowledge of the angles of the **PSG** components, and the angles and precise properties of aluminium turning mirrors. The performance of GASP in the laboratory using this matrix is given below.

For any astronomical polarimeter calibration is an essential and difficult process. A general method for determining the system matrix in the laboratory is described above. However, for the telescope as a whole to be a complete polarimeter the system matrix must include any polarisation effects in the atmosphere and the telescope, which is more difficult to achieve. Only the instrument itself can be calibrated and to polarimetrically calibrate the telescope and instrument, four linearly independent Stokes vectors must be injected into the input *before* the telescope pupil - which is nearly impossible to implement. Likewise, if GASP is to be used in an off-axis focus, such as a Nasmyth focus, this will have a considerable influence upon **A**, mainly by introducing unwanted changes of polarisation to the beam. However, in general if the folding mirror has no multilayer dielectric coatings (which could produce nearly 100% attenuation of certain polarisation states) it will still be possible to invert **A**, although the condition number will be lowered and systematic errors can bias the estimation. Knowing the complex refractive index of the mirror a theoretical Mueller matrix for the folding mirror can be derived (Van Harten et al. 2009) and used to determine the Stokes vector before being reflected and altered by the mirror.

In practice, calibration sources will be built into GASP, and (hopefully) small corrections to the measured Stokes vectors made by observations of known "linear polarisation calibration" stars (Fossati et. al 2007). This can then correct, as an offset, the change in degree of polarisation and induced circular polarisation caused by the telescope and confirm the rotational alignment of the instrument while on the telescope.

**Polarimetric Verification Tests:**

Laboratory tests of GASP were conducted at a wavelength of 650 nm, using filtered (15nm bandwidth) white light from an incandescent source. Tests were conducted in two modes, a "linear test" whereby a polariser was rotated in front of the source and a "circular test" whereby a QWP followed the rotating polariser yielding polarisation states going from left-hand to right-hand circularly polarised, through 100% linear polarisation. Figure 7 shows, for input 100% linear polarised light at various angles, the measured degrees of total polarisation (DoP), linear polarisation (DoLP), and circular polarisation (DoCP) as a function of the input angle. At all angles the precise vales are 100% for DoP and DoLP, and 0% for DoCP. The residuals (figure 8) show a systematic error and an average offset of ~0.7% for DoLP and ~1.35% for DoCP.

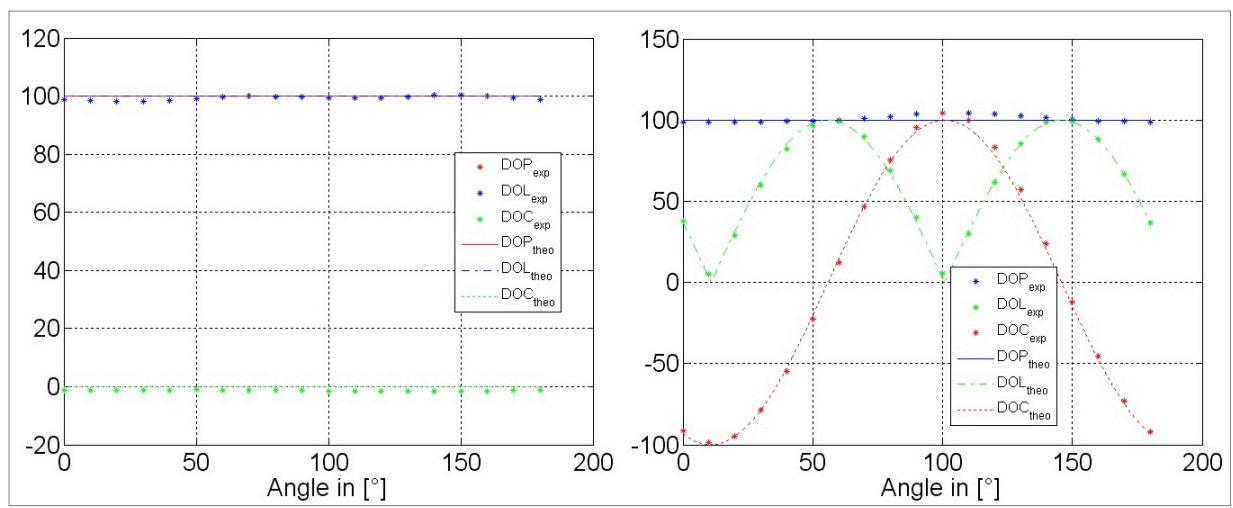

**Fig. 7** Verification test for the measurement of degrees of polarisation for the "linear" test (left panel), and "circular test" (right panel), produced when a quarter wave plate is inserted following the rotating 100% linear polariser.

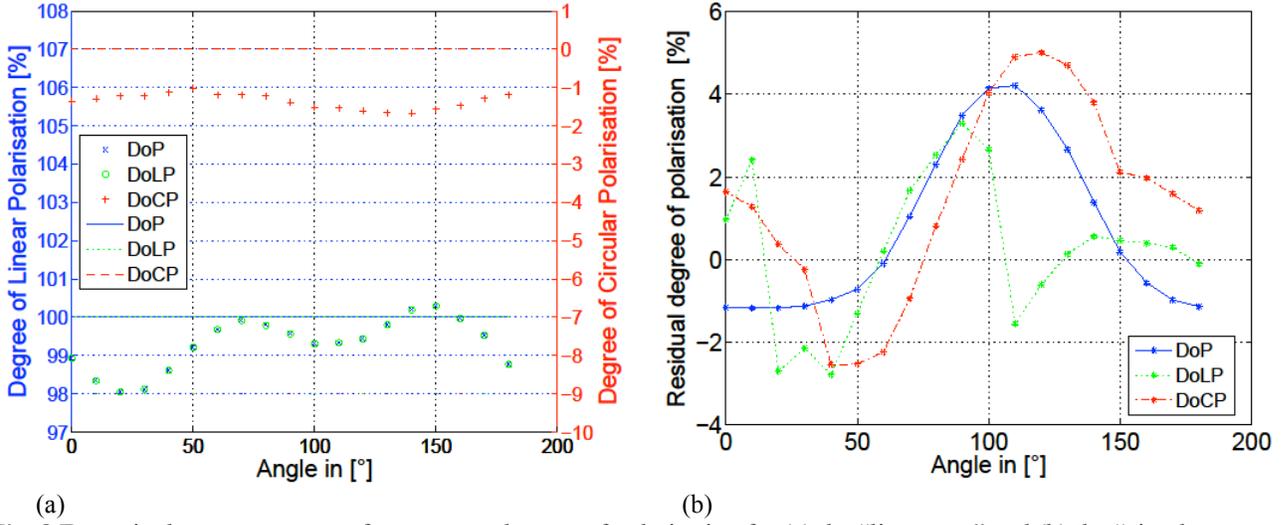

(a)               (b)

**Fig. 8** Errors in the measurement of percentage degrees of polarisation for (a) the "linear test" and (b) the "circular test". In each case the "angle in" refers to the angle of the rotating linear polariser. Expected degrees of polarisation are shown in figure 7. Small offset (~1%) and systematically varying errors (<2°) are observed in laboratory tests. Rapid changes in the residual error for the DoLP for the "circular test" at ~10° and ~100°, occur when the value of DoLP becomes very small.

GASP measures the angles of linear polarisation as well as the degrees of polarisation. Figure 9, below, shows a comparison between the expected and measured angles for the "linear test" and "circular test", which again are very close to the expected angles. Figure 10, below gives the values of the residual difference between these angles, again as a function of linear polariser angle.

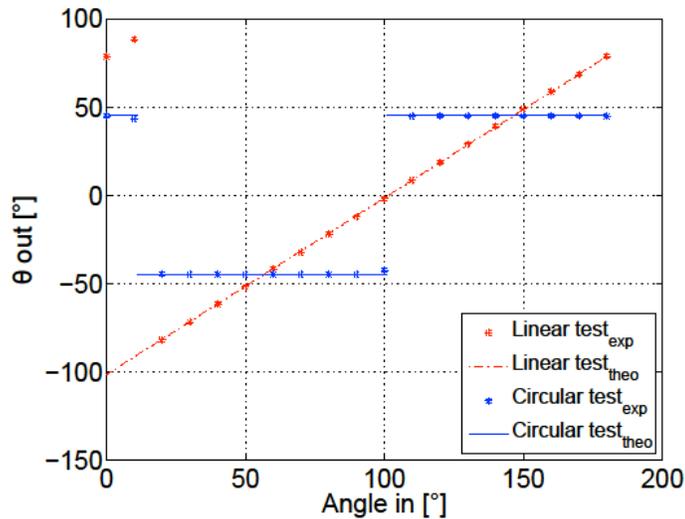

**Fig. 9** Polarimetric verification test for the output angle of linear polarisation (θ). Both the "linear test" and "circular test" are presented as measured by GASP (see paragraph for explanation). Both tests show that the theoretical and experimental results are a good match. The shift at the start of the "linear test" is due to the fact that GASP outputs θ as -90° to 90° only. Note that when the linear and circular polarisers produce 100% circular polarisation (at 10° and 100° in the diagram) the angle of linear polarisation is undefined.

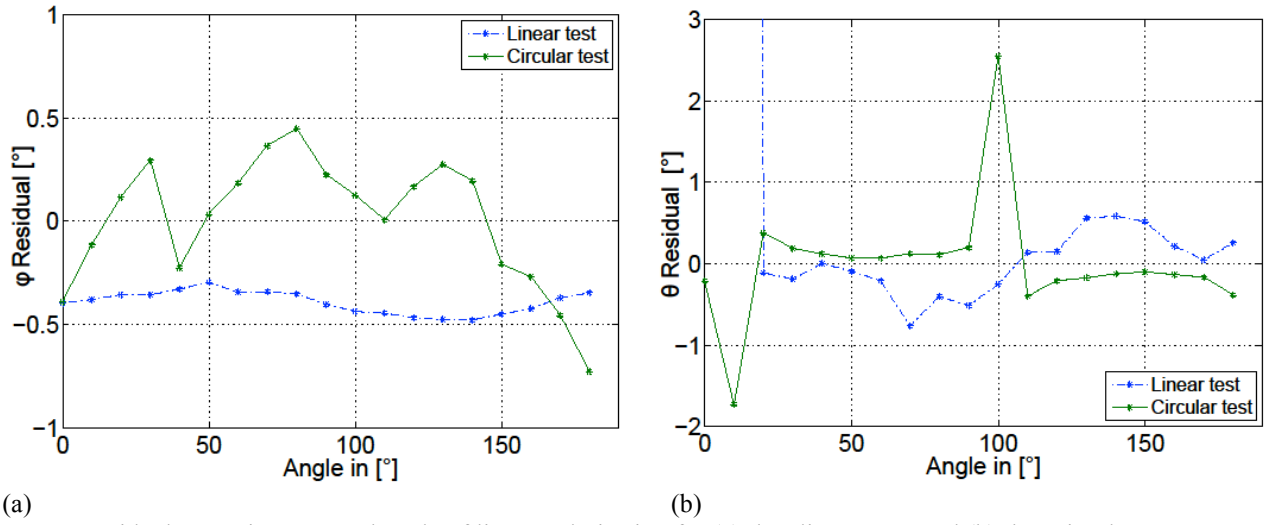

(a)                                            (b)

**Fig.10** Residual errors in measured angle of linear polarisation for (a) the "linear test" and (b) the "circular test" as a function of the input linear polariser angle. A small offset (<0.5°) a systematically varying (<0.5°) error can be seen. The large errors at ~10° and ~100° correspond to polariser positions where the input beam becomes ~100% circularly polarised – and the angle of linear polarisations becomes undefined.

## 4. Noise Sensitivity of GASP

A photon counting polarimeter like GASP – which relies upon comparing the intensities of four light beams - is inherently subject to Poisson noise at various stages in its calibration and use. The effect of noise in calibration is to produce (fixed) errors in the detector calibration (**A**) matrix which lead to small systematic errors in the measurement of the components of the Stokes vector of the incoming light, but it does not introduce random noise. These systematic effects during calibration can be minimised by combining very large numbers of counts at each calibration angle with the averaging effects of either of several calibration methods. One may use arbitrarily large numbers of counts during calibration, subject only to detector count rate and detector linearity limitations, and/or to detector gain stability considerations if large numbers of counts are accumulated over a long calibration time.

The effect of noise in actual target measurements is more complex, affecting different Stokes vectors differently. This is because certain extreme values – for example nearly 100% polarised light, either linear or circular, at particular orientations – could lead to one of the output intensities $i_1$, $i_2$, $i_3$ or $i_4$ becoming very small. Then small statistical fluctuations in one of these numbers become exaggerated in the mathematics of the matrix inversion used to generate the input Stokes vector from the output intensity vector. In the case of astronomical targets, however, these extreme situations essentially never occur and for the purposes of investigating the Poisson-noise sensitivity of GASP, polarised light with 20% linear and 0% circular polarisation, and a total count rate of $3.10^6$ counts per second subject to Poisson fluctuations, has been simulated and then analysed. The results of one such simulation run are given in figure11, below.

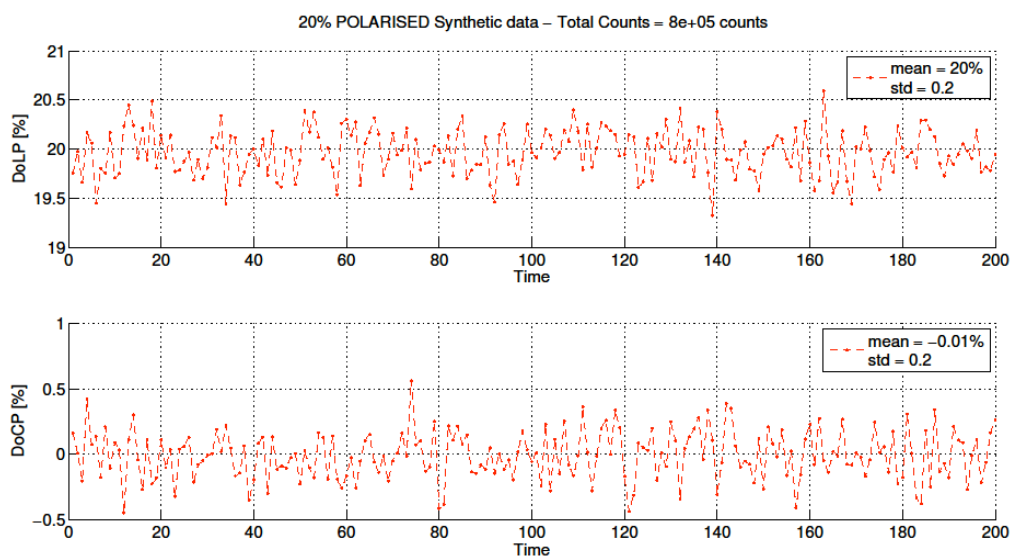

**Fig. 11** Monte Carlo Simulation of GASP data using a theoretical calibration matrix to model both the ingoing light and outgoing polarisation values (Degree of Linear and Degree of Circular polarisation). The simulation was modeled for a wavelength of 589nm (for the prism glass type).

Extensive simulation illustrated in Figure 11 show that, in the situation where the counts are approximately equally distributed between the four output channels, output noise in the measurement of the degree of linear polarisation (DoLP) has a noise corresponding to Poisson fluctuations in 50% of the total count, and in the degree of circular polarisation (DoCP) to 100% of the total count. This enables us to compute the sensitivity of GASP as a function of bandwidth for different brightness targets observed on different sized telescopes. Detailed investigations of the noise sensitivity of GASP will be the subject of a later paper.

## 5. Preliminary on-sky calibrations and preliminary results

GASP was used on the 1.2m Cassini Telescope at Loiano Observatory on 29.08.2008 to observe the highly polarised source CRL2668 (Michalsky et al 1976). CRL2668 is a compact reflection (possibly proto-planetary) nebula. It has three approximately equally spaced sources in a line of length ~20 arcseconds (therefore spatially resolvable by GASP), with V magnitudes 14.0, 12.7, and 16.7 (Turnshek et al 1990). The brightest (centre) source has linear polarisation ~50% and circular polarisation ~-0.6%.It is not expected to show any degree of rapid variability, and so is a suitable calibration source for GASP. The observed counting rate was ~$50 \times 10^3$ cps., corresponding to a statistical fluctuation level of 0.2% in the degree of linear polarisation in a one second integration, according to our simulations.

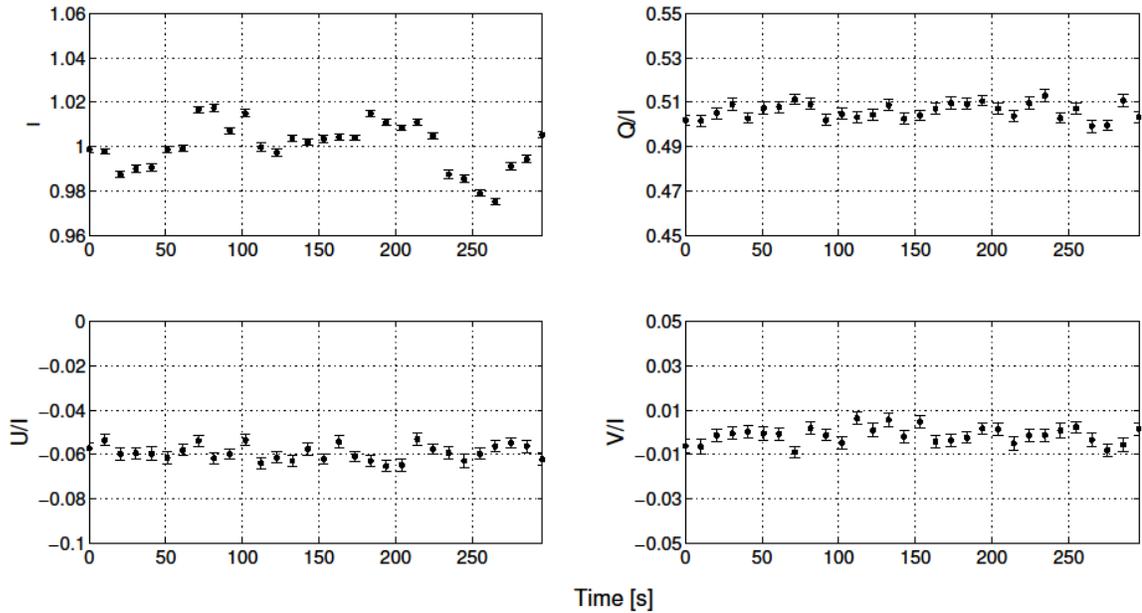

**Fig. 12** Temporal Stability of Observed Stokes parameters of CRL2688 with GASP, normalized to the value of I to take into account variable seeing and extinction. Counts are binned into 10 second intervals. Error bars are plotted for the standard error ($\sigma$) of each individual data point. Variability in the measured values for the whole data set are: Q/I (1.5$\sigma$), U/I (1.4$\sigma$), and V/I (1.4$\sigma$).

Figure 12 shows the Stokes parameters for CRL2688, normalized to the value of I, with the data binned into 10s intervals. During this time the seeing was quite variable, however, there is only a small degree of variability greater than purely statistical fluctuations over the 300s interval. Hence variations in degree of polarisation of ~0.2% could be detected for a suitably bright source. However, *systematic uncertainties in the calibration process lead to systematic errors which can be much larger than this*. This issue will be addressed in section 6, below.

The measured degrees of polarisation were: linear polarisation 51.0% (± 0.04% statistical precision), and circular polarisation -0.1% (± 0.05% statistical precision). The overall intensity was highly variable due to effects like seeing fluctuations and atmospheric transparency fluctuations. Nevertheless, measurements of the degrees of polarisation display no fluctuations which can be obviously attributed to these effects. This is one of the major strengths of the division of amplitude polarimeter, which makes it so valuable for rapidly varying sources, particularly in poor conditions.

Systematic effects during calibration are observed in our data, in, for example, the absolute angles of the linear polariser and quarter wave retarder, and/or imperfections in the polarising components themselves, giving rise to errors in the coefficients of the system matrix, or, indeed, to linear polarisation imparted by the telescope optics. For this reason *the absolute values of the degrees of polarisation cannot be determined with this data set to within the statistical precision, but temporal variations of the order of the statistical precision could be determined*. This issue will be addressed in section 6, below.

## 6. Discussion and Conclusion

### 6.1 Performance of GASP

GASP has been demonstrated both in the laboratory and on the sky to be capable of simultaneous measurement of the four components of the full Stokes vector for astronomical sources. The particular strength of the design of GASP is the ability to perform these measurements under less than perfect conditions (of seeing, telescope tracking etc.), and to obtain measurements having fluctuations limited primarily by statistical fluctuations in the numbers of photons counted in the four channels.

Based upon the laboratory and on-sky calibration GASP has been determined to be capable of measuring the degrees of linear and circular polarisation to 0.5%, and variability in these to 0.2%, for a sufficiently bright source. The time resolution of GASP is determined (a) by the time resolution of the detector(s) used, and (b) by statistical fluctuations in the signals. For CRL2688, using a 1.5m telescope, these figures correspond to a 10 second integration.

However, it has been much more difficult to obtain an absolute calibration consistent with the degree of statistical precision that can be obtained. There are several reasons for this. Firstly, whilst measurements of the four output components are made simultaneously, calibration measurements (using at least four input Stokes vectors, with, for example, 100% linear polarisation at various angles and 100% circular polarisation) must be made sequentially – so any systematic variations (in detector gain, for example) will produce errors in the A matrix and bias in the polarimeter calibration. Secondly, nearly perfect optical components, required for pseudo-inverse calibration, may be difficult to obtain, as is error-free component absolute alignment. We are therefore currently working on the implementation of the Eigenvalue Calibration Method (ECM) by Compain, Poirier & Drevillon (1999)to eliminate all linear systematic variations in the system.

The ECM was originally designed to operate on Mueller matrix polarimeters. It can reliably isolate errors associated with the Polarisation State Analyzer (PSA, GASP in this case) and Polarisation State Generator (PSG, the calibration optics in this case) independently. At first sight, GASP may not seem to be a Mueller matrix polarimeter, however it is a full Stokes vector polarimeter. Our current PSG (used in the pseudo-inverse calibration) can generate linearly independent states of polarisation, therefore when the PSA in GASP and the PSG are used together, the instrument is, in fact, a complete Mueller matrix polarimeter. The ECM can lead to high degrees of repeatability and accuracy (Lara and Dainty 2006), and for this reason we have been adjusting our data processing to calibrate GASP using this method.

The principle of the ECM is rather simple, and thorough explanation is readily available in the literature mentioned above. The essence of the method is that four polarisation samples can be used to perform the calibration, and that the exact properties of these samples can be derived from the eigenvalues of the non-calibrated matrices as initially measured with the instrument. The samples are aligned, one at a time, between the PSG and the PSA – this is what mathematically allows for the separation of the errors in the PSG from the errors in the PSA. Ultimately, during our observations, we want to correct only the errors in the PSA, but it is paramount that we know if there are any errors in the PSG to evaluate our PSA correctly.

### 6.2 Potential Targets for GASP

GASP is designed to measure both linear and circular optical polarisation on short times scales for both periodic and randomly fluctuating sources. Three primary classes of targets have been initially identified as listed below, which require the unique capability of GASP. Our theoretical estimates are based upon using GASP mounted on a 4-m class telescope using sCMOS detector technology under moderate seeing conditions of 2 arcseconds.

1. Measurements of the optical polarisation during randomly-occurring giant radio pulses (GRP) for the Crab pulsar, for which GASP would be sensitive to variations in the polarisation of the order of 0.1% between normal and GRP events. In making this estimate we have assumed an observation of 5 hours, with a GRP rate of 0.5 Hz identified by simultaneous radio observations.

2. Measurements of the optical polarisation of magnetars or anomalous X-ray pulsars, which would be sensitive to measuring polarisation at the 8% level for a 26th magnitude target from 10 hours of observation.

3. Measurement of the optical polarisation of the optical after glow from gamma ray bursts. Assuming a 22nd magnitude afterglow. GASP can measure polarisation at the 1% level on a 100 second time scale.

As well as these targets GASP can be used in the study of a wide range of astronomical objects including accreting systems onto magnetized white dwarfs and neutron stars; flare stars; brown dwarfs and changes in polarisation during exo-planet transits.

## Acknowledgements

The authors wish to thank Science Foundation Ireland for funding the development of GASP under grant number 09/RFP/AST239. We are grateful to Loiano Observatory (Osservatorio Astronomico di Bologna) and to Calar Alto Observatory (Centro Astronómico Hispano Alemán) for provision of telescope time during the development and commissioning of GASP.

The authors also wish to thank the referees/reviewers of the draft of this paper for their detailed and helpful comments.

## Acknowledgements

The authors wish to thank Science Foundation Ireland for funding the development of GASP under grant number 09/RFP/AST239. We are grateful to Loiano Observatory (Osservatorio Astronomico di Bologna) and to Calar Alto Observatory (Centro Astronómico Hispano Alemán) for provision of telescope time during the development and commissioning of GASP.

The authors also wish to thank the referees/reviewers of the draft of this paper for their detailed and helpful comments.